# A New Design for Array Multiplier with Trade off in Power and Area.

N. Ravi[1], A. Satish[2], Dr. T. Jayachandra Prasad[3] and Dr. T. Subba Rao[4]

[1] Research Scholar(SVU), Department of Physics,
[2, 3] Department of ECE,
[1, 2, 3] RGMCET (Autonomous), JNT University-Anantapur, Nandyal, AP-518501, India.
[1]*ravi2728@gmail.com.*
[2]*sathish_anchula@yahoo.com.*
[3]*jp.talari@gmail.com.*

[4] Department of Physics, S.K.University,
Anantapur, AP-515003, India.
*sushmasurekha@yahoo.com.*

## Abstract

In this paper a low power and low area array multiplier with carry save adder is proposed. The proposed adder eliminates the final addition stage of the multiplier than the conventional parallel array multiplier. The conventional and proposed multiplier both are synthesized with 16-T full adder. Among Transmission Gate, Transmission Function Adder, 14-T, 16-T full adder shows energy efficiency. In the proposed 4x4 multiplier to add carry bits with out using Ripple Carry Adder (RCA) in the final stage, the carries given to the input of the next left column input. Due to this the proposed multiplier shows 56 less transistor count, then cause trade off in power and area. The proposed multiplier has shown 13.91% less power, 34.09% more speed and 59.91% less energy consumption for TSMC 0.18nm technology at a supply voltage 2.0V than the conventional multiplier.

## 1. Introduction

Multiplication is an essential arithmetic operation for common Digital Signal Processing (DSP) applications, such as filtering and fast Fourier transform (FFT). To achieve high execution speed, parallel array multipliers are widely used. But these multipliers consume more power. Power consumption has become a critical concern in today's VLSI system design. Hence the designers are needed to concentrate power efficient multipliers for the design of low-power DSP systems.

In recent years, several power reduction techniques have been proposed for low-power digital design, including the reduction of supply voltage, multi threshold logic and clock speed, the use of signed magnitude arithmetic and differential data encoding, the parallelization or pipelining of operations, and the tuning of input bit-patterns to reduce switching activity [1].

A basic multiplier can be divided into three parts i) partial product generation ii) partial product addition and iii) final addition. In this paper we present a low power design methodology for parallel array multiplier using Carry Save Adder (CSA). The rest of this paper is organized as follows. Section-II presents the total power consumption in CMOS circuits with mathematical expression. Section-III explains the basic structure of an array multiplier with mathematical expression. The methodology of the proposed multiplier with conventional array multiplier is presented in Section-IV. Results of total power, worse case delay and EDP with different technologies is discussed in Section-V. Section-VI is the conclusion of the work.

## 2. Power Consumption in CMOS VLSI Circuits

There are three main components of power consumption in digital CMOS VLSI circuits.

**2.1) *Switching Power*:** consumed in charging and discharging of the circuit capacitances during transistor switching.

**2.2) *Short-Circuit Power*:** consumed due to short-circuit current flowing from power supply to ground during transistor switching. This power more dominates in Deep Sub Micron (DSM) technology.

**2.3) *Static Power*:** consumed due to static and leakage currents flowing while the circuit is in a stable state. The first two components are referred to as dynamic power, since power is consumed dynamically while the circuit is changing states. Dynamic power accounts for the majority of the total power consumption in digital CMOS VLSI circuits at micron technology [2], [3].

$$P = \sum i V_{DD} V_{swing} C_{load} f P i + V_{DD} \sum_i I_{isc} + V_{DD} I_l$$

-------- (1)

Where
$V_{DD}$ – power supply voltage;

$V_{swing}$ - voltage swing of the output which is ideally equal to $V_{DD}$;
$C_{load}$ – load capacitance at node i;
f – system clock frequency;
$P_i$ – switching activity at node I;
$I_{isc}$ - short-circuit current at node;
$I_l$ – leakage current.

As designing a low power CMOS 1-bit full adder, the emphasis will be on these areas

i) to reduce the total number of transistors and the total number of parasitic capacitances in internal nodes to reduce the load capacitance.

ii) to lower the switching activity to save the dynamic power consumption.

iii) to remove some direct paths from power supply to ground to save the short-circuit power dissipation.

iv) to balance each path in the full adder to avoid the appearance of glitches since glitches not only cause a unnecessary power dissipation hut may even lead to a fault circuit operation due to spurious transitions, especially in a low voltage operation system.

v) in order to build a low-voltage full adder, all the nodes in the circuit must possess full voltage swing.

vi) to build the low-voltage full adder design because the power supply voltage is the crucial factor in reducing power dissipation.

In Nanometer scale leakage power dominates the dynamic power and static power due to hot electrons. So the concentration is on to trade off power in parallel multipliers.

## III. PARALLEL MULTIPLIER:

Consider the multiplication of two unsigned n-bit numbers, where $X = x_{n-1}, x_{n-2}, \ldots x_0$ is the multiplicand and $Y = y_{n-1}, y_{n-2}, \ldots y_0$ is the multiplier. The product of these two bits can be written as [4]

$$P = \sum_{i=1}^{n-1} X_i \sum_{j=1}^{n-1} Y_j 2^{(i+j)} \quad \text{---------- (2)}$$

$$X = \sum_{i=1}^{n-1} X_i 2^i \quad \text{----- Multiplicand}$$

$$Y = \sum_{j=1}^{n-1} Y_j 2^j \quad \text{------ Multiplier}$$

In the given example, we have 4-bit multiplier and 4-bit multiplicand. By using the above equation (2) we can generate 4-rows of partial products as shown in the Fig (1).The hardware required for the generation of these partial products is AND gates. Using any adder like Carry Save Adder (CSA), Carry Propagate Adder (CPA) we can add the partial products. In this method we are following Carry Save Addition to add the products.

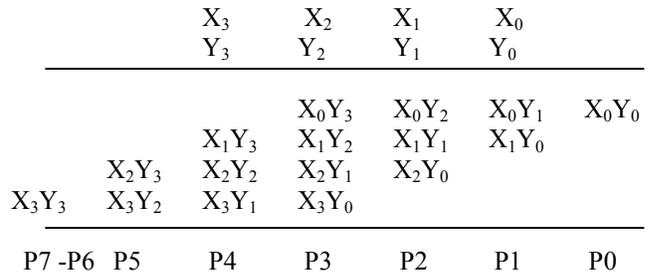

|  |  |  |  | $X_3$ | $X_2$ | $X_1$ | $X_0$ |
|---|---|---|---|---|---|---|---|
|  |  |  |  | $Y_3$ | $Y_2$ | $Y_1$ | $Y_0$ |
|  |  |  | $X_0Y_3$ | $X_0Y_2$ | $X_0Y_1$ | $X_0Y_0$ |
|  |  | $X_1Y_3$ | $X_1Y_2$ | $X_1Y_1$ | $X_1Y_0$ |  |
|  | $X_2Y_3$ | $X_2Y_2$ | $X_2Y_1$ | $X_2Y_0$ |  |  |
| $X_3Y_3$ | $X_3Y_2$ | $X_3Y_1$ | $X_3Y_0$ |  |  |  |
| P7 -P6 | P5 | P4 | P3 | P2 | P1 | P0 |

Fig.1. Multiplier Architecture.

## 3. Methodology

In the Carry Save Addition method, the first row will be either Half-Adders or Full-Adders. If the first row of the partial products is implemented with Full-Adders, Cin will be considered '0'. Then the carries of each Full-Adder can be diagonally forwarded to the next row of the adder. The resulting multiplier is said to be Carry Save Multiplier, because the carry bits are not immediately added, but rather are saved for the next stage. In the design if the full adders have two input data the third input is considered as zero. In the final stage, carries and sums are merged in a fast carry-propagate (e.g. ripple carry or carry-look ahead) adder stage [5]. This is the conventional array multiplier with CSA as shown in "Fig. (2)".

In the proposed method, we implement all the partial product rows of the multiplier as same as that of the conventional adder (explained above).The final adder which is used to add carries and sums of the multiplier is removed in this method. Then the carries of the multiplier at the final stage is carefully added to the inputs of the multiplier as shown in the "Fig (3)". The carry of the fourth column of the multiplier is given to the input of the fifth column instead of zero. Then the carry of the fifth column is forwarded to the input of the sixth column so on. In this multiplier the carry of the seventh column of the adder is not neglected, it is considered as Most Significant Bit (MSB) of the multiplier. Due to elimination of four full adders in the final stage power and area can be trade off in the proposed design than that of the conventional array multiplier.

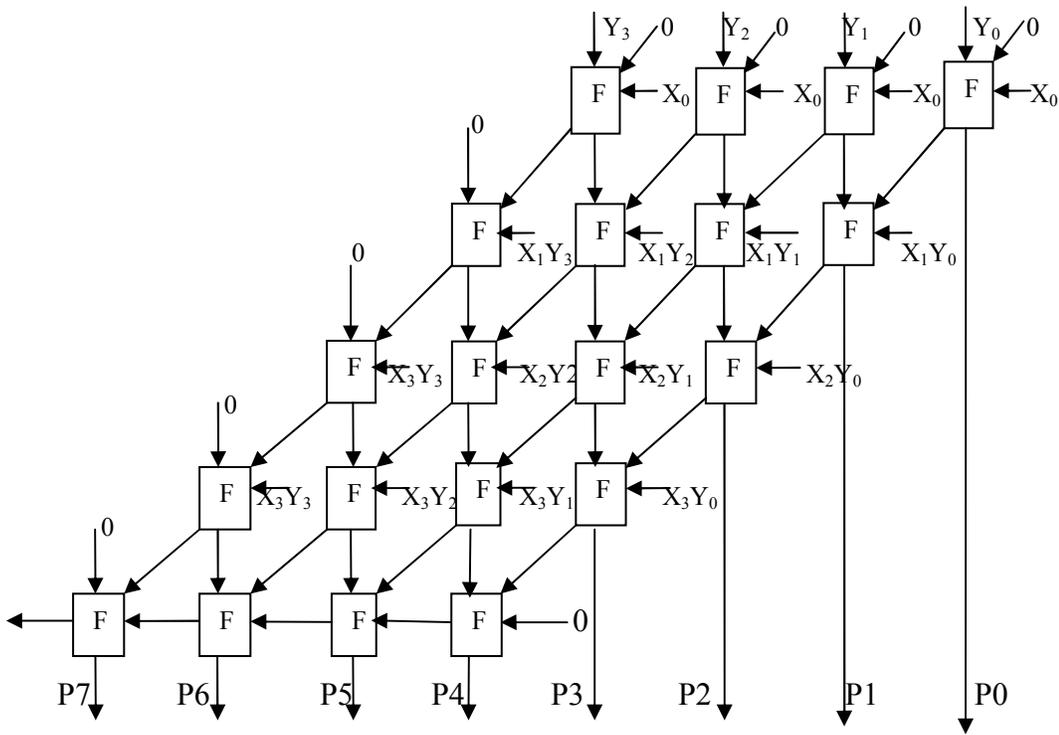

Fig. 2. Conventional Array Multiplier with CSA.

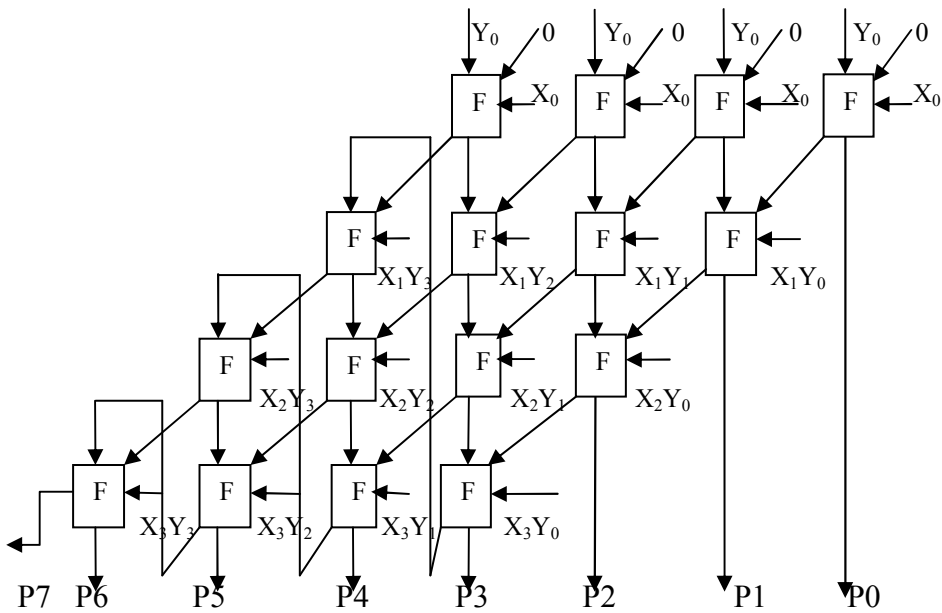

Fig .3. Proposed Array Multiplier with CSA.

## 4. Results and Discussions

To show the proposed design very good performance, the design is synthesized with 16-Transistor Full Adder design. Among the Transmission Gate CMOS, Transmission Function Full Adder (TFA), 14-T adders this shows good efficiency in terms of power and delay [6]. The results of the 16-T full adder are as shown in the "Table. (1)".

Table.1. Power, Delay and Energy Delay Product of the 16-T Adder.

| Technology at ($25^0$C) | 16-T Full-Total Power | 16-T Full-Prop-Delay | EDP (JS) |
|---|---|---|---|
| 0.18um | 8.88E-06 | 5.08E-10 | 2.29161E-24 |
| 90nm | 1.36E-05 | 5.07E-10 | 3.49587E-24 |
| 65nm | 6.15E-06 | 5.06E-10 | 1.57462E-24 |

Power, Propagation delays and Energy consumptions are calculated for the conventional array and the proposed multiplier with different technologies using H-Spice to study the performance of the both multipliers shown in table (2).

- **4. (i) Total Power:** The total power of these two multipliers is calculated at a simulation temperature of $25^0$C as shown in "Fig. 4". The proposed multiplier has shown the power improvement than the conventional for 180nm - 13.91 %, 90nm – 13.71% and 65nm – 18.59%. In the DSM technology the proposed multiplier show good performance due to less transistor count to avoid more leakage current.

- **4. (ii) Propagation Delay:** In the propagation delays of the two multipliers, the proposed has shown more efficient than the convention shown in "Fig. 5". The propagation delay is calculated for all inputs and outputs to find the worst case delay.

- **4. (iii) Energy Delay Product:** The proposed multiplier shows more energy efficiency is as shown in "Fig. 6". It has shown for 180nm – 59.91%, 90nm – 9.35% and 65nm – 29.21% improvement in the energy.

- **4. (iv) Transistor count:** The proposed multiplier has 56 less transistors than that of the conventional multiplier. These less count can trade off power consumption and area.

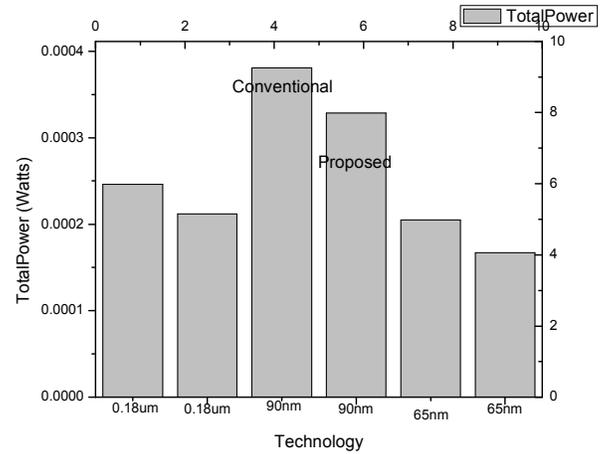

Fig. 4. Total Power Comparison.

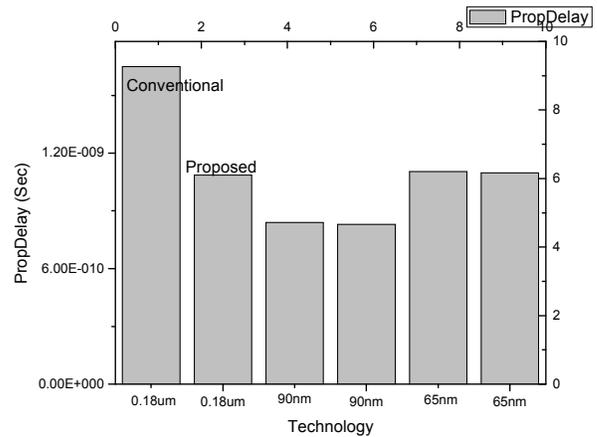

Fig. 5. Delay Graph.

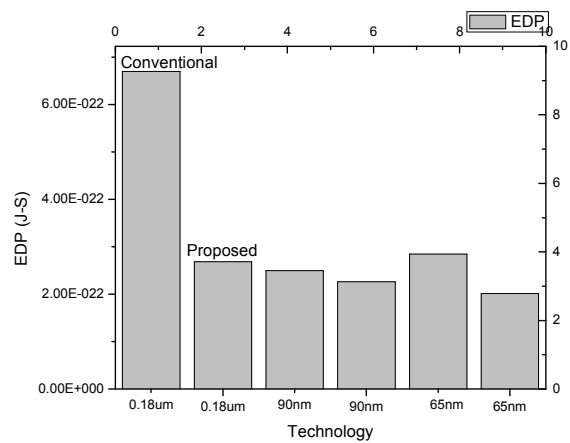

Fig. 6. EDP Graph.

Table. 2. Power, Delay and Energy Delay Product of the Conventional and Proposed Multiplier.

| Technology | Array Multiplier type | Total Power (Watts) | Total Power Percentage (%) | Prop-Delay (Sec) | Prop-Delay Percentage (%) | Energy Delay Product (EDP) JS | EDP Percentage (%) | No.of Transistors |
|---|---|---|---|---|---|---|---|---|
| 0.18um | Conventional | 2.4628E-04 | 13.91 | 1.6490E-09 | 34.09 | 6.6968E-22 | 59.91 | 376 (Conventioal) |
|  | Proposed | 2.1200E-04 |  | 1.0867E-09 |  | 2.6841E-22 |  |  |
| 90nm | Conventional | 3.8089E-04 | 13.71 | 8.3947E-10 | 1.12 | 2.5002E-22 | 9.35 |  |
|  | Proposed | 3.2864E-04 |  | 8.3000E-10 |  | 2.2664E-22 |  | 320 (Proposed) |
| 65nm | Conventional | 2.0514E-04 | 18.59 | 1.1040E-09 | 0.52 | 2.8451E-22 | 29.21 |  |
|  | Proposed | 1.6699E-04 |  | 1.0982E-09 |  | 2.0139E-22 |  |  |

## 5. Conclusion

In this paper we have proposed a new design for low power, high performance and low area based array multiplier with out RCA. It shows the same functionality than the conventional adder. For higher bit multiplication it shows better power and area saving. For example, in the proposed 4x4 multiplier it saves 56 transistors. For TSMC 0.18um it saves 13.91% of total power, 34.09% of more speed and 59.91% less energy consumption. To study the performance of the multiplier, it is synthesized with different technologies.

**N.Ravi** holds B.Sc degree in electronics at Osmania Degree College-Kurnool, S.K. University, Anantapur, AP-India in 1998. He obtained Masters Degree in Physics (Solid State Physics) at S.K.University, Anantapur, AP- India in 2001. He is pursuing Ph.D in VLSI Design from Sri Venkateswara University, Tirupati. He is presently working at Rajeev Gandhi memorial College of Engg & Technology, Nandyal, Andhra Pradesh-India as an Assistant professor in the department of Physics. His area of interest includes VLSI, Solid State Physics and Nanotechnology.

**Dr.T.Jayachandra Prasad** is a Principal and Professor of ECE at Rajeev Gandhi Memorial College of Engg & Technology, Nandyal, Andhra Padesh-India

**Dr.T.Subba Rao** is a Professor in the department of Physics, Sri Krishnadevaraya University, Anantapur, Andhra Pradesh-India.